\documentclass[11pt,a4paper]{article}
\usepackage{jheppub}

\newcommand {\dd}{\mathrm{d}}
\usepackage{epsfig,multicol,bbm,amsxtra,subfigure,ulem}
\title{ Localization of $U(1)$ gauge vector field on  flat branes with { five}-dimension (asymptotic) AdS$_{5}$ spacetime }

\author[a]{Zhen-Hua Zhao\footnote{Corresponding author},}
\author[b,c]{Qun-Ying Xie}

\affiliation[a]{
    Department of Applied Physics,
    Shandong University of Science and Technology,
    Qingdao, 266590 People's Republic of China}
\affiliation[b]{School of Information Science and Engineering,
Lanzhou University, Lanzhou 730000, People's Republic of China }
\affiliation[c]{Institute of Theoretical Physics,
        Lanzhou University, Lanzhou 730000,
        People's Republic of China
    }
\emailAdd{zhaozhh78@sdust.edu.cn}
\emailAdd{xieqy@lzu.edu.cn}

\abstract{

In order to localize $U(1)$ gauge vector field on Randall-Sundrum-like braneworld model with infinite extra dimension, we propose a new kind of non-minimal coupling between the $U(1)$ gauge field and the gravity. We propose three kinds of coupling methods and they all support the localization of zero mode. In addition, one of them can support the localization of massive modes. Moreover, the massive tachyonic modes can be excluded. And our method can be used not only in the thin braneword models but also in the thick ones.

}


\begin{document}
\maketitle

\section{Introduction}
In the braneworld theory our universe can be taken as a brane embedding in the higher dimensional spacetime. So the localization of matter fields is an very important issue. Needless appending extra terms in  the standard action { the} gravity can be localized on branes \cite{Randall199983,Randall199983a,Gremm2000478} as well as scalar fields \cite{Bajc2000}. 
The localization of { fermions} can be realized  by introducing the usual Yukawa  coupling \cite{JackiwRebbi1976} (or by another form of coupling \cite{LiuXuChenWei2013}) with background scalar field  in thick brane models \cite{Rubakov1983,Randjbar-Daemi2000,Ringeval200265,Koley200522,Melfo2006,LiuZhangZhangDuan2008,Liu2018}. 
The localization of $U(1)$ gauge field is more complex than above two fields, { especially,} in Randall-Sundrum  (RSII)  like model \cite{Randall199983a}. Pomarol showed in Ref. \cite{Pomarol2000} that $U(1)$ gauge field can not be localized in the RSII model with the following action 
\begin{equation}
S=-\frac{1}{4}\int \dd^5x\sqrt{-g}F_{MN}F^{MN}, \label{action0}
\end{equation}
where $F_{MN}=\partial_M A_N-\partial_N A_M$ is the field strength.
But in some non-RSII models with non-flat brane $U(1)$ fields 
can be localized, for examples,  the thick de Sitter (dS) brane models \cite{Liu20090902,GuoHerrera-AguilarLiuMalagon-MorejonMora-Luna2013,Herrera-AguilarRojasSantos-Rodriguez2014},  the thick Weyl  brane model \cite{Liu200808}, and  the thin wave brane models \cite{GogberashviliSingleton2010,GogberashviliMidodashviliMidodashvili2012}.

{ The aim of this paper is to argue a new method to localize $U(1)$ gauge field in 5D RSII-like braneworld models. For a RSII-like braneworld scenario which should hold the $Z_{2}$ symmetric along the infinite extr-dimension,} 5D spacetime is {an} (asymptotic) anti-de Sitter (AdS$_{5}$) { one}, and the brane is  flat. There are three common methods to realize the localization of U(1) gauge field.

The first { one } is to multiply $F_{MN}F^{MN}$ by a dilaton factor. For { examples,} in the work of Ref.~\cite{Kehagias2001504}, dilaton factor is a function of scalar dilaton field and in the work of Refs.~\cite{ChumbesHoffHott2012,SuiZhao2017} the dilaton factor is a function of the background scalar field.  It should be noted that  the above two works only  hold in thick brane models, because the scalar fields they needed only appears in thick brane models.

The second one is to append a mass term to above standard action ~\eqref{action0}. For examples, in the work of Ref.~\cite{Ghoroku2002} the authors appended a bulk constant mass term and a boundary mass term into the action, and in the work of Ref.~\cite{ZhaoXieZhong2014} the authors took the non-minimally coupling of vector and gravity into account,  where the 5D scalar curvature is taken as the mass of 5D vector. {What should  be noted is  the scalar curvature to be taken as the mass of 4D  vector is first  proposed in Ref.~\cite{GolovnevMukhanovVanchurin2008}  to be used in  
studying the inflation of our 4D universe. } 
 But adding a mass term into  the 5D vector fields action will break its gauge invariance, so in Ref.~\cite{Vaquera-AraujoCorradini2015} the authors proposed a 5D Stueckelberg-like action to correct this problem.

 The third one is to add a topology term to the 5D vector action. In Ref.~\cite{Oda2001} Oda proposed a method by { adding} a 3-form potential and { a} topological term  to the 5D vector field action to realized the localization of zero mode.


%
%

In this { work} we will follow the idea of the first approach \cite{Kehagias2001504,ChumbesHoffHott2012}. But in our method the dilaton factor is not a function of scalar fields but a function of scalar curvature. So  our method can be used  not only to thick brane models but also to the thin branes models.

This paper is constructed as follows. We  review our method in section \ref{LocGF}. The localization of zero and massive modes are discussed in section \ref{sec3} and  section \ref{sec4}, respectively. Applying  our method  to a concrete  braneworld model is shown in section \ref{sec5}.  Finally, we give our conclusions in section \ref{Cons}.

\section{The {  Method} }\label{LocGF}

 The branes discussed here are  flat and the line element of the five-dimensional space-time is assumed to be 
\begin{eqnarray}
\dd s^2={g}_{MN}\dd x^Mdx^N=e^{2\alpha(y)}\eta_{\mu\nu}\dd x^{\mu}\dd x^{\nu}+\dd y^2,\label{metric}
\end{eqnarray}
where $e^{2 \alpha(y)}$ is the warp factor, $\alpha(y)$ is only the function of extra-dimensional coordinate $y$, and $\eta_{\mu\nu}=\text{diag}{(-1,1,1,1)}$ is the metric on the branes.
The braneworld  which  holds an  (asymptotic) AdS$_{5}$ spacetime at infinity means that the scalar curvature $R$ satisfies 
\begin{equation}
\lim_{y\to \pm \infty}R(y)=-{\mathrm{C}_{R}},\label{Cond1}
\end{equation}
where $\mathrm{C}_{R}>0$ is a constant.
By using of the metric \eqref{metric} one can get the expression of $R$:
\begin{eqnarray}
R(y)=-20 \alpha'(y)^{2}-8 \alpha''(y). \label{Ry}
\end{eqnarray}
To be consistent with the condition (\ref{Cond1}), $\alpha(y)$ { should have} the following asymptotic solution
{ 
\begin{equation}
\alpha({ y \to \pm \infty} )\to  \mp \,k\, y + \mathrm{C}.\label{alphainf}
\end{equation}
}
where parameter $k>0$ and
 $ \mathrm{C} $ is a constant, without loss of generality we set $\mathrm{C}=0$ here and after.
 {  Substituting \eqref{alphainf} into \eqref{Ry} one can get 
\begin{equation}
{\mathrm{C}_{R}}=20 k^2.
\end{equation}
}

The action of the five-dimensional $U(1)$ gauge field we proposed is
\begin{equation}
S=-\frac{1}{4}\int \dd^5 x \sqrt{-g}F(R) F_{MN}F^{MN}\label{action4},
\end{equation}
where $F(R)$ is a function of scalar curvature $R$, against the action in Ref.~\cite{ZhaoXieZhong2014}, this action is gauge invariant.

How to confirm the form of $F(R)$? The rules {  we follow} are:
\begin{enumerate}
\item when $R \to 0$ the action should return to the standard form (\ref{action0});
\item {  $F(R) $  should satisfy  the positivity condition 
\begin{equation}
  F(R) >0 \label{positivity}
\end{equation}
and the finity condition at the same time }
\begin{equation}
\int_{-\infty}^{+\infty}  F(R)\, \dd y  < \infty \label{finity}
\end{equation}
 to preserve the canonical form of 4D action.
\end{enumerate}
  {  Models  discussed here hold $\mathbb{Z}_2$} symmetrical along the extra-dimension, so $\alpha(y)$ is an even function. 
{ 
If $R(y)$ is monotonously decreasing as  $y$ varying form 0 to $\pm \infty$,  $F(R)$ can  has a simple form
\begin{equation}
F(R)=\chi(R)=1+\frac{R}{20 k^2 }\label{FR1}
\end{equation}
}
to satisfy the  positivity condition  \eqref{positivity}. 
If {  $R(y)$} varies non-monotonic along the extra-dimension the form of $F(R)$ in \eqref{FR1} can not  guarantee  {  the positivity condition  \eqref{positivity}. But} a form of  { 
\begin{equation}
F(R)=\frac{1}{N}\sum_{ n=1}^{N}(\chi(R)^2)^{n},\label{FR2}
\end{equation}
 where $N>0$ and  $n$ is  an integer},  can hold the positivity condition \eqref{positivity}, {  obviously}.

It can be  proven that  the functions of $F(R)$ in equations \eqref{FR1} and \eqref{FR2} can only help us to obtain the localized zero mode. In order to  obtain localized massive modes, we  propose the following form:
{ 
\begin{equation}
F(R)=e^{\mathrm{C}_{2}\left(1-\left(\chi(R)^2\right)^{-\mathrm{C}_{3}/2} \right)},\label{FR3}
\end{equation} 
}
where $\mathrm{C}_{2}$ and $\mathrm{C}_{3}$ are under-determined positive parameters. {  We will show  $\mathrm{C}_{3}$ is related to the asymptotic behavior of the potential at infinity in eq. \eqref{Vz}.}


\section{Localization of the Zero Mode}\label{sec3}

By means of the decomposition  $A_{\mu}=\sum_n a_{\mu}(x)\rho_n(y)$ and the gauge condition $\partial_{\mu} A^{\mu}=0$ and $A_4=0$, the above action (\ref{action4}) is  reduced to
\begin{equation}
S=-\frac{1}{4}\int \dd y F(R) \rho_n(y)^2\int \dd^4x(f_{\mu\nu}f^{\mu\nu}-2 m_n^2 a_{\mu}a^{\mu})\label{action5},
\end{equation}
where $f_{\mu\nu}=\partial_{\mu}a_{\nu}-\partial_{\nu}a_{\mu}$ is the four-dimensional gauge field strength tensor, and $\rho_n(y)$  satisfies the equation
\begin{equation}
\rho_n''+\left(\frac{F'}{F}+2\alpha'\right)\rho_n'=-m_n^2\rho_n e^{-2\alpha} \label{eq},
\end{equation}
{  with the  boundary conditions either the Neumann $\rho_n'(\pm \infty)=0$ or the Dirichlet $\rho_n(\pm \infty)=0$ \cite{Gherghetta2010},  }  where the prime `` $'$ '' stands for the derivative with respect to $y$ in this section.
 The localization of gauge field requires
\begin{equation}
I\equiv\int_{-\infty}^{+\infty} \dd y F(R)\rho_n^2(y)=1 \label{int}.
\end{equation}

For the zero mode, $m_0=0$, eq. (\ref{eq}) reads
\begin{equation}
\rho_0''+\left(\frac{F'}{F}+2\alpha'\right)\rho_0'=0. \label{eq0}
\end{equation}
{  
By setting  $ \gamma'=2\alpha' + F'/F $,  the above equation \eqref{eq0} reads
\begin{equation}
\rho_0''+\gamma' \rho_0'=0. 
\end{equation} 
The general solution of zero mode is
\begin{equation}
\rho_0=\mathrm{c}_0 +\mathrm{c}_1 \int e^{-\gamma} \dd y. \label{appzeroso1}
\end{equation} 
Here $\mathrm{c}_0$ and $\mathrm{c}_1$ are arbitrary constants.
 Models  discussed here hold the $\mathbb{Z}_2$ symmetry along the extra-dimension, $F(R)$, $\alpha(y)$ and  $\gamma$ are even functions  of $y$, so  the second term in eq. \eqref{appzeroso1} is odd. The Dirichlet boundary conditions  $\rho_n(\pm \infty)=0$ will lead to $\mathrm{c}_0=0$ and $\mathrm{c}_1=0$. But the Neumann boundary  conditions $\rho_n'(\pm \infty)=0$ only lead  to $\mathrm{c}_1=0$, so the zero mode solution is
 \begin{equation}
\rho_0=\mathrm{c}_0 .
\end{equation} 
} 
The localization {  about zero mode}	 can be realized  when  $F(R)$ satisfies the finity condition \eqref{finity},  namely
\begin{eqnarray}
& &\int_{-\infty}^{+\infty} \dd y F(R)\rho_0^2(y)\nonumber\\
&=&\mathrm{c}_0 ^{2}\int_{-\infty}^{+\infty}  F(R)\, \dd y=1 \label{int2}.
\end{eqnarray}
Because $F(R)$ is continuous, the convergence of the above integration is determined  by the asymptotic behavior of $F(R)$ at  infinity along the extra dimension. The convergent condition is
\begin{equation}
F(R(y\to \pm \infty)) \propto  y^{-p} \label{Cd2}
\end{equation}
 with $p > 1$.
{ 
By using the $\mathbb{Z}_2$ symmetry,  we may only need to discuss the case of $y\to + \infty$.  The way of $\alpha(y)$ to its limit solution  \eqref{alphainf} is  model dependent.} In order to discuss the asymptotic behavior of $\chi$ at infinity, we add the first order infinitesimal term into $\alpha(y)$ \eqref{alphainf}
{ 
\begin{equation}
\alpha(y \to + \infty) \sim - \,k\, y - \mathrm{C} y^{-q}.\label{alphainf2}
\end{equation}
where $q >0$. In order to satisfy positive condition  $\mathrm{C}>0$ is need for $F(R)$ in eq. \eqref{FR1}. 
Substituting \eqref{alphainf2}  and \eqref{Ry} into \eqref{FR1} }  one get 
\begin{equation}
\chi(R(y \to + \infty)) \sim \frac{2 \mathrm{C} q\, y^{-(q+1)}}{k} \label{chiinf}
\end{equation}
by neglecting  higher order infinitesimal terms. It is easy to test that under the above asymptotic solution, the integration of $F(R)$, with the forms of \eqref{FR1}, \eqref{FR2} and \eqref{FR3},  are convergent. This leads to the conclusion that the zero mode can be localized.


\section{Localization of Massive Modes} \label{sec4}


To discuss the localization of massive modes,
it is more convenient to do the following coordinate transformation
\begin{equation}\label{ytoz}
\left\{
\begin{array}{lll}
\dd z &=& e^{- \alpha(y)} \dd y\\
z     &=& \int  e^{- \alpha(y)} \dd y 
\end{array} \right.
\end{equation}
with the boundary condition $z(y=0)=0$. The line element \eqref{metric} is recasted to the following conformally flat form
namely,
\begin{equation}
\dd s^2=e^{2 \alpha(z)}(\eta_{\mu\nu}\dd x^{\mu}\dd x^{\nu}+\dd z^2).\label{gcf}
\end{equation}
Further,  by using of the gauge choice $A_4=0$ and the decomposition  $A_{\mu}=\sum_n a_{\mu}(x)\rho_n(z)$ the action of the five-dimensional gauge field (\ref{action4})
is reduced to
\begin{equation}
S=-\frac{1}{4}\sum_{n}\int \dd z \tilde{\rho}^2_n(z)\int \dd^4x(f^{(n)}_{\mu\nu}f^{(n)\mu\nu}-2 m_n^2 a^{(n)}_{\mu}a^{{(n)}\mu})\label{action7},
\end{equation}
{ 
where $\tilde{\rho}_n=\rho_n \sqrt{F(R)} e^{\alpha/2}$, 
}
and $\tilde{\rho}_n(z)$ satisfies the following Schr\"odinger-like equation
\begin{equation}
-\tilde{\rho}''_n+V(z)\tilde{\rho}_n= m_n^2\tilde{\rho}_n, \label{eq3}
\end{equation}
where the effective potential is given by
\begin{equation}
V(z)=\frac{1}{2}\alpha''(z)+\frac{1}{4}
   \alpha'^2(z)+\frac{\alpha'(z) F'(R)}{2 F(R)}+\frac{F''(R)}{2 F(R)}-\frac{F'^2(R)}{4 F^2(R)}.\label{Vz}
\end{equation}
In this section the prime denotes the derivative with respect to the coordinate $z$.
Equation \eqref{eq3} can be recast to
\begin{equation}
\bigg(\frac{\dd}{\dd z}+\Gamma'(z)\bigg)\bigg(-\frac{\dd}{\dd z}+\Gamma'(z)\bigg)\tilde{\rho}_n=m_n^2\tilde{\rho}_n, \label{eq4}
\end{equation}
where
\begin{equation}
\Gamma'(z)=\frac{1}{2}\left(\frac{F'(R)}{F(R)}+\alpha'(z)\right),
\end{equation}
and without loss of the generality we can take
\begin{equation}
\Gamma(z)=\frac{1}{2}\left(\ln{F(R)}+\alpha(z)\right).
\end{equation}
Equation \eqref{eq4} means that there is no tachyonic mode with $m_{n}^2<0$ in the spectrum of the KK modes \cite{Bazeia2004}.
 It is worth to note that in order to get the effective action (\ref{action7}) of the four-dimensional  vector field from the five-dimensional one (\ref{action4}), we have introduced the orthonormalization condition between different massive  modes:
\begin{equation}
\int \dd z \tilde{\rho}_m(z) \tilde{\rho}_n(z) = 0. ~~~(m\neq n) 
\end{equation}
So the localization condition for $\tilde{\rho}_n(z)$ is
\begin{equation}
\int \dd z \tilde{\rho}^2_n(z)<\infty.
\end{equation}
The solution of zero mass mode at the $z$ coordinate is easy to be found from \eqref{eq4},
{ 
\begin{equation}
\tilde{\rho}_0(z)=e^{\alpha(z)/2}\sqrt{F(R)}. \label{zeroz1}
\end{equation}
}

The existence of massive modes is determined by the potential $V(z)$. Based on the condition \eqref{alphainf} we have shown the existence of zero mode, this means that  at  some finite area $V(z)<0$ along the extra-dimension. So a necessary condition of the existence of  localized massive modes is  $V(\pm\infty)>0$.   {   $V(\pm\infty)$  is determined  by asymptotic behavior of  $\alpha(z)$ and  $F(R(z))$.  In the following we will discuss the limit of $V(z)$ with $z \to + \infty$ only, based on the $\mathbb{Z}_2$ symmetry  $V(-\infty)$ is the same to $V(+\infty)$. }

By using the {  limit solution} $\alpha(y)$ \eqref{alphainf}  and the coordinate transformation  \eqref{ytoz},
we obtain the  {  limit solution} of $\alpha$ with $z$ coordinate at infinity,
\begin{equation}
 \alpha(z)\to -\ln ( k\, z).\label{alphainfz}
\end{equation}
{  For different models the ways to this limit solution are different. In order to discuss the asymptotic behavior of $V(z)$ at infinity, we add a first order infinitesimal term to $ \alpha(z)$, namely, 
\begin{equation}
 \alpha(z)\to -\ln ( k\, z)+\mathrm{C}_4 z^{-n},      \label{alphainfzasy} 
\end{equation}
where $\mathrm{C}_{4}$ and $n$ are model dependent constants and $n>0$. 
The scalar curvature $R$ at the $z$ coordinate reads
\begin{equation}
 R(z) = -4 e^{2 \alpha(z)} (3\alpha'(z)+2 \alpha'' (z)).\label{fRz}
\end{equation}
Substituting \eqref{alphainfzasy} into \eqref{fRz} one can get the asymptotic solution of $R$,
\begin{eqnarray}
 R(z \to + \infty) \to && -20 k^2  e^{-2\mathrm{C}4 z^{-n}} \nonumber  \\
 && - (8 \mathrm{C}_4 k^2 n^2 +32 \mathrm{C}_4 k^2 n)   e^{-2 \mathrm{C}_4 z^{-n}} z^{-n} \nonumber \\
  && -12 \mathrm{C}_4^2 k^2 n^2  e^{-2 \mathrm{C}_4 z^{-n}}  z^{-2 n}
 . \label{fRzinf1} 
\end{eqnarray}
Because 
\begin{equation}
\lim_{z \to +\infty} e^{-2 \mathrm{C}_4 z^{-n}}=1,
\end{equation}
 and by ignoring the $z^{-2n}$ term, 
we can further reduce $ R(z \to + \infty) $ into 
\begin{equation}
 R(z \to +\infty )\to -20 k^2 + \mathrm{C}_5 \, z^{-n}, \label{Rzinf} 
\end{equation}
where $\mathrm{C}_{5}=- (8 \mathrm{C}_4 k^2 n^2 +32 \mathrm{C}_4 k^2 n) $.

 Substituting eq. \eqref{Rzinf} into eq. \eqref{FR1}, one can get the   asymptotic solutions of $\chi(R)$, namely,
 
  \begin{eqnarray}
\chi(R(z \to +\infty))  \to  \frac{\mathrm{C}_5 \, z^{-n}}{20 k^2}.  \label{chiz}
 \end{eqnarray}
 
For the first case of $F(R)$ in   eq. \eqref{FR1}, 
substituting eqs.~\eqref{chiz} and \eqref{alphainfz}  into eq.~\eqref{Vz}, we obtain the asymptotic solution of $V(z)$,
 \begin{eqnarray}
V_{\mathrm{I}}(z \to +\infty) \to \frac{1}{4} \mathrm{C}_4^2 n^2 z^{-2 n-2}+\mathrm{C}_4 n (n+1) z^{-n-2}+\frac{n^2+4 n+3}{4 } z^{-2}.
 \end{eqnarray}
 
 For the second case of  $F(R)$ in eq. \eqref{FR2},   the  asymptotic solution of $V(z)$ is
 \begin{eqnarray}
V_{\mathrm{II}}(z \to +\infty) \to (n^2+2 n+\frac{3}{4} )z^{-2} ,
 \end{eqnarray}
 by ignoring the higher order infinitesimal terms. 
 The limits  of $V_{\mathrm{I}}(z) $ and $V_{\mathrm{II}}(z) $ are
  \begin{eqnarray}
\lim_{z\to +\infty} V_{\mathrm{I}}(z)=\lim_{z\to +\infty} V_{\mathrm{II}}(z) = 0, \label{VI1}
 \end{eqnarray}
 so there are no massive mode in that two cases.

For the third case of $F(R)$ in eq. \eqref{FR3}, the asymptotic solution of $V(z)$ is
\begin{eqnarray}
V_{\mathrm{III}}(z \to +\infty)&&\to
\mathrm{C}_6 z^{2 \mathrm{C}_{3} n -2}+\mathrm{C}_7 z^{\mathrm{C}_{3} n -2}+\mathrm{C}_8 z^{\mathrm{C}_{3} n- n-2}\nonumber\\
&&+\frac{1}{4} \mathrm{C}_{4} ^2 n^2 z^{-2 (n+1)}+\frac{1}{2} \mathrm{C}_{4}  n^2 z^{- n-2}+\mathrm{C}_{4}  n z^{- n-1}+\frac{3}{4} z^{-2}, 
\end{eqnarray}
where 
\begin{eqnarray}
\mathrm{C}_6&=&
\mathrm{C}_{2} ^2 4^{2 \mathrm{C}_{3} -1} 25^{\mathrm{C}_{3} } \mathrm{C}_{3} ^2 n^2 |\mathrm{C}_{5}|  ^{-2 \mathrm{C}_{3} } k^{4 \mathrm{C}_{3} },\nonumber\\
\mathrm{C}_7 &=&( 2 - \mathrm{C}_{3}  n ) \mathrm{C}_{2} 2^{2 \mathrm{C}_{3}-1 } \mathrm{C}_{3}  n  5^{\mathrm{C}_{3} } |\mathrm{C}_{5}|  ^{-\mathrm{C}_{3} } k^{2 \mathrm{C}_{3} },
 \nonumber\\
\mathrm{C}_8 &=&\mathrm{C}_{2}  2^{2 \mathrm{C}_{3} -1} 5^{\mathrm{C}_{3} } \mathrm{C}_{3}  \mathrm{C}_{4}  n^2 |\mathrm{C}_{5}|  ^{-\mathrm{C}_{3} } k^{2 \mathrm{C}_{3} } .
\nonumber
\end{eqnarray}
From the above asymptotic solution, we find }
\begin{equation}\label{Vzinf2}
 V_{\mathrm{III}}(\infty)=\left\{
\begin{array}{llcll}
+\infty         & &\mathrm{C}_{3} > {1}/{n}   & & \text{infinitely deep potential}\\
\mathrm{C}        & &\mathrm{C}_{3} = {1}/{n}   & & \text{P\"{o}schl-Teller potential}\\
0                 & & 0 <\mathrm{C}_{3} < {1}/{n}& & \text{volcanic potential}
\end{array} \right. 
\end{equation}
where $\mathrm{C}$ is a positive constant.

So  the condition $\mathrm{C}_{3} = 1/n$ means a finite number of localized massive modes, when $\mathrm{C}_{3} > 1/n$, there will be  infinite number of localized massive modes, and $ 0 <\mathrm{C}_{3} < {1}/{n}  $ corresponds to  no massive mode can be localized on branes.

\section{Concrete  Braneworld  Model} 
\label{sec5}

{  In this section only  the first case \eqref{FR1} and third case \eqref{FR3}  of $F(R)$ are discussed, since the conclusion about second case \eqref{FR2} of  $F(R)$  is the same with the first one  \eqref{FR1}.  }  The model {  we used } is proposed in Ref. ~\cite{ZhongLiu2016} without scalar field, where the five-dimensional action is
\begin{equation}
S=\frac{1}{4} \int \dd ^{4} x \dd y \sqrt{- g} f(R),
\end{equation} 
and $f(R)$ reads
\begin{eqnarray}
\label{SolF1}
f(R)&=&\frac{4}{7} \left(6 k^2+R\right)\cosh (\beta(w(R)) )
\nonumber\\
&-&\frac{2}{7} k^2 \sqrt{480-\frac{36 R}{k^2}-\frac{3 R^2}{k^4}} \sinh (\beta(w(R)) ),
\end{eqnarray}
where $\beta (w)$ is defined as
\begin{eqnarray}
\beta (w)\equiv 2 \sqrt{3} \arctan \left(\tanh \left(\frac{w}{2} \right)\right),
\end{eqnarray}
with
\begin{eqnarray}
\label{WR}
w(R)=\pm\text{arcsech}\left[\frac{\sqrt{20 +R/k^2}}{2\sqrt{7}}\right].
\end{eqnarray}
With above conditions the solution of $\alpha(y)$ is 
\begin{equation}
\alpha(y)= - \ln (\cosh (k\, y)).
\end{equation}
Under the $z$ coordinate  
\begin{eqnarray}
\alpha(z)&=&-\frac{1}{2} \log \left(k^2 z^2+1\right), \label{chiexam1}
\end{eqnarray}
and
%
{ 
\begin{eqnarray}
R(z)&=&-4 e^{-2 {\alpha}(z)} \left(2 {\alpha}''(z)+3 {\alpha}'(z)^2\right)\nonumber\\
    &=&\frac{8 k^2-20 k^4 z^2}{k^2 z^2+1}. \label{Rzexam1}
\end{eqnarray}
With the above analytic solutions of $\alpha(z)$ \eqref{chiexam1} and $R(z)$ \eqref{Rzexam1}, one can easily get 
\begin{eqnarray}\label{Vz1}
V_{\mathrm{I}}(z)=\frac{3 k^2 (-2 + 5 k^2 z^2)}{4 (1 + k^2 z^2)^2},
\end{eqnarray}
and }
\begin{eqnarray}\label{Vz1}
{  V_{\mathrm{III}}(z)     }
&=&\frac{1}{4} \left(\frac{25}{49}\right)^{\mathrm{C}_{3}} k^2 \left(\frac{1}{k^2 z^2+1}\right)^{2-2 \mathrm{C}_{3}} \bigg[4 \mathrm{C}_{2}^2 \mathrm{C}_{3}^2 k^2 z^2 +49^{\mathrm{C}_{3}} \left(3 k^2 z^2-2\right) \left(\frac{1}{5 k^2 z^2+5}\right)^{2 \mathrm{C}_{3}}  \nonumber\\
& &-4 \mathrm{C}_{2} 7^{\mathrm{C}_{3}} \mathrm{C}_{3} \left(2 (\mathrm{C}_{3}-1) k^2 z^2+1\right) \left(\frac{1}{5 k^2 z^2+5}\right)^{\mathrm{C}_{3}}\bigg].
\end{eqnarray}
{ 
The limit of $ V_{\mathrm{I}}(z)$ is
\begin{eqnarray}
\lim_{z\to +\infty} V_{\mathrm{I}}(z) = 0,  \label{VI2}
 \end{eqnarray} 
that  is consistent with eq. \eqref{VI1}. }
The asymptotic behavior of $R(z)$ is 
\begin{eqnarray}
R(z \to \infty )=-20 k^{2}+8 z^{-2}. \label{Rzinf2}
\end{eqnarray}
Compare  eq. \eqref{Rzinf2}  with  eq.  \eqref{Rzinf}, one can find that
\[ n=2, \]
which can be verified by the limit of $V_\mathrm{III}(z)$ \eqref{Vz1},  namely,
\begin{equation}
V_{\mathrm{III}}(\pm\infty)=\left\{
\begin{array}{llcll}
+\infty  & &\mathrm{C}_{3} > 1/2 , \\
{5 \mathrm{C}_{2}^2 k^2}/{28} & & \mathrm{C}_{3} =1/2 ,  \\
0        & &0 <\mathrm{C}_{3} < 1/2,
\end{array} \right.  \label{Vzinf3}
\end{equation}
{  Equations  \eqref{Vzinf3} and \eqref{VI2} can be taken as a test of our conclusion made in  section \ref{sec4}. }

\begin{figure}[htbp]
\begin{center}
\includegraphics[width= 0.49\textwidth]{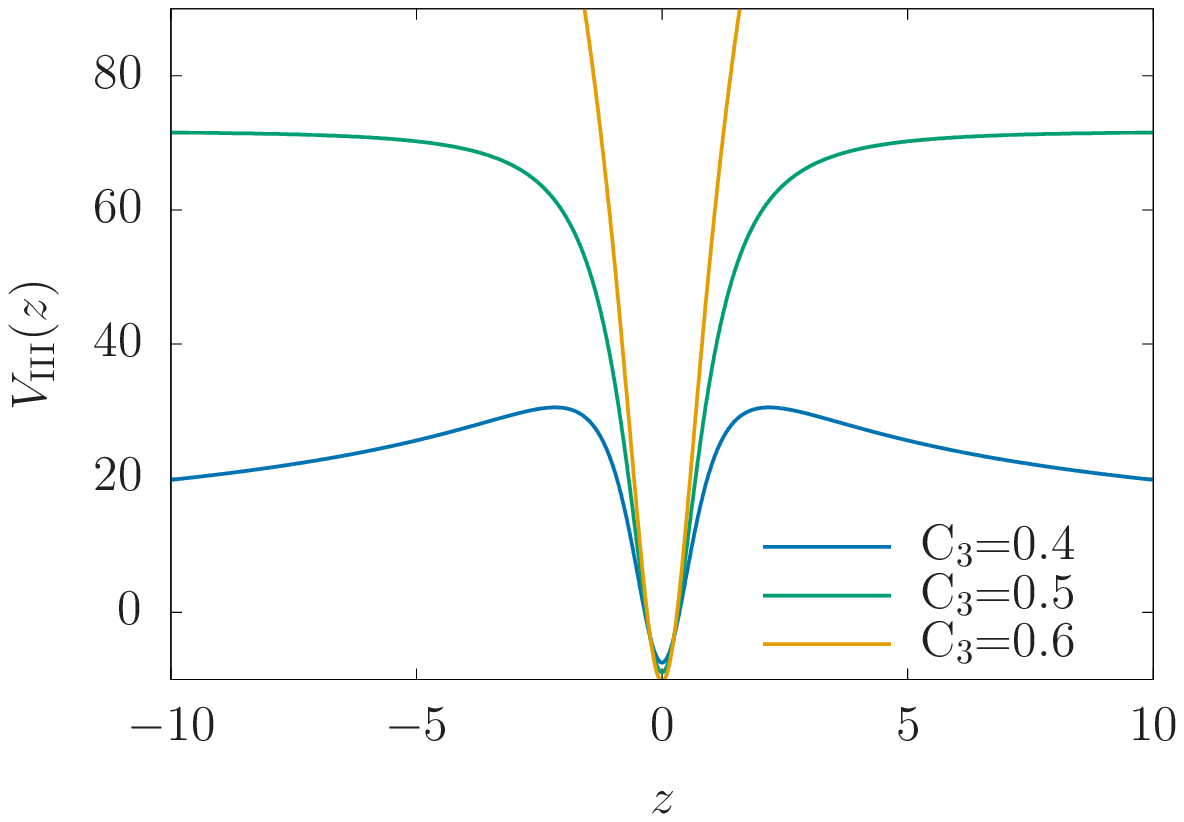}
\includegraphics[width= 0.49\textwidth]{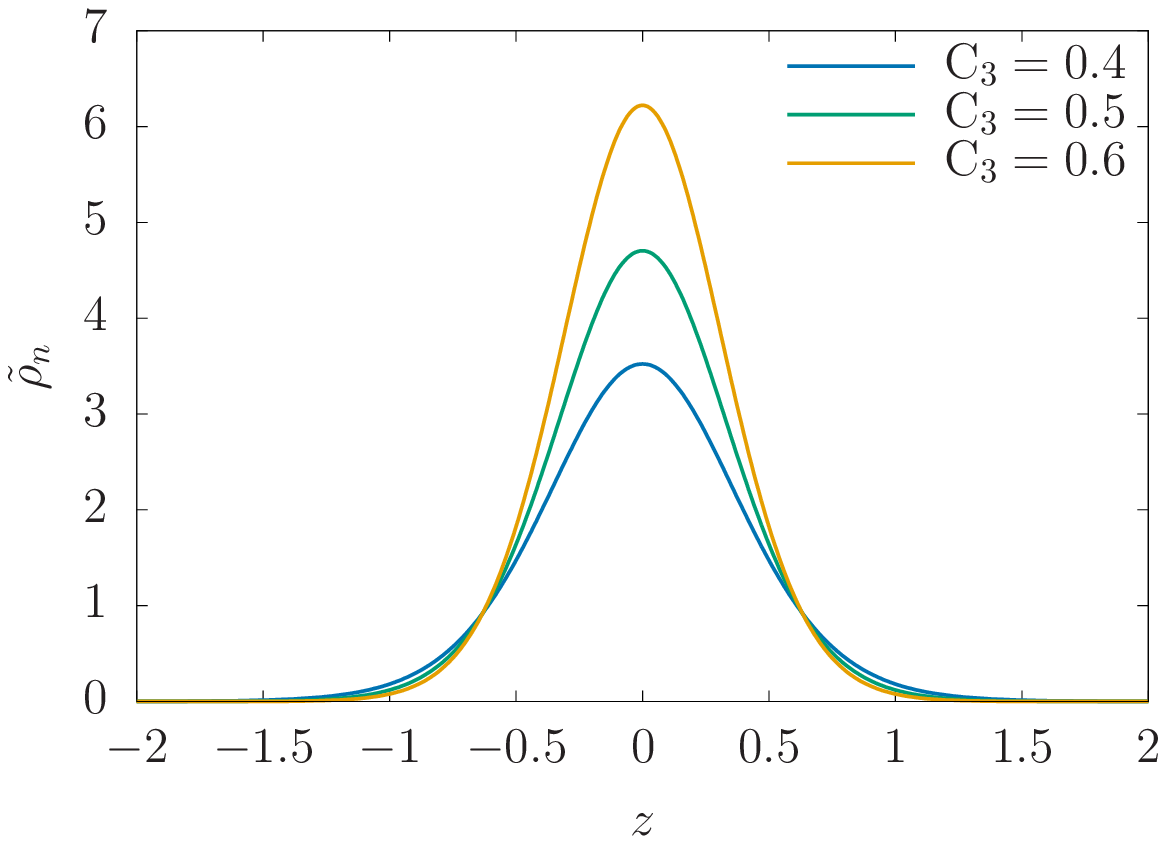}
\caption{Plots of $V_\mathrm{III}(z)$ and $\tilde{\rho}_{0}$  with $\mathrm{C}_{2}=20,\;k=1.$}
\label{figVz1rhoz1}
\end{center}
\end{figure}

{  In the following we show some mode solutions with the potential $V_\mathrm{III}(z)$ \eqref{Vz1}. The zero mode solution \eqref{zeroz1} reads
 \begin{equation}
 \tilde{\rho}_{0}(z)={\sqrt{e^{\mathrm{C}_{2}-\mathrm{C}_{2} \left(\frac{5}{7}\right)^{\mathrm{C}_{3}} \left(\frac{1}{k^2 z^2+1}\right)^{-\mathrm{C}_{3}}}}}{{(k^2 z^2+1)}^{-1/4}}.
 \end{equation}
 }
The plots of $V_\mathrm{III}(z)$ and  $\tilde{\rho}_{0}(z)$ { (non-normalized) with} some specific values of parameters of $\mathrm{C}_{2},\;k,$ and $\mathrm{C}_{3}$ are shown in figure \ref{figVz1rhoz1}.

\begin{figure}[htbp]
\begin{center}
\includegraphics[width= 0.49\textwidth]{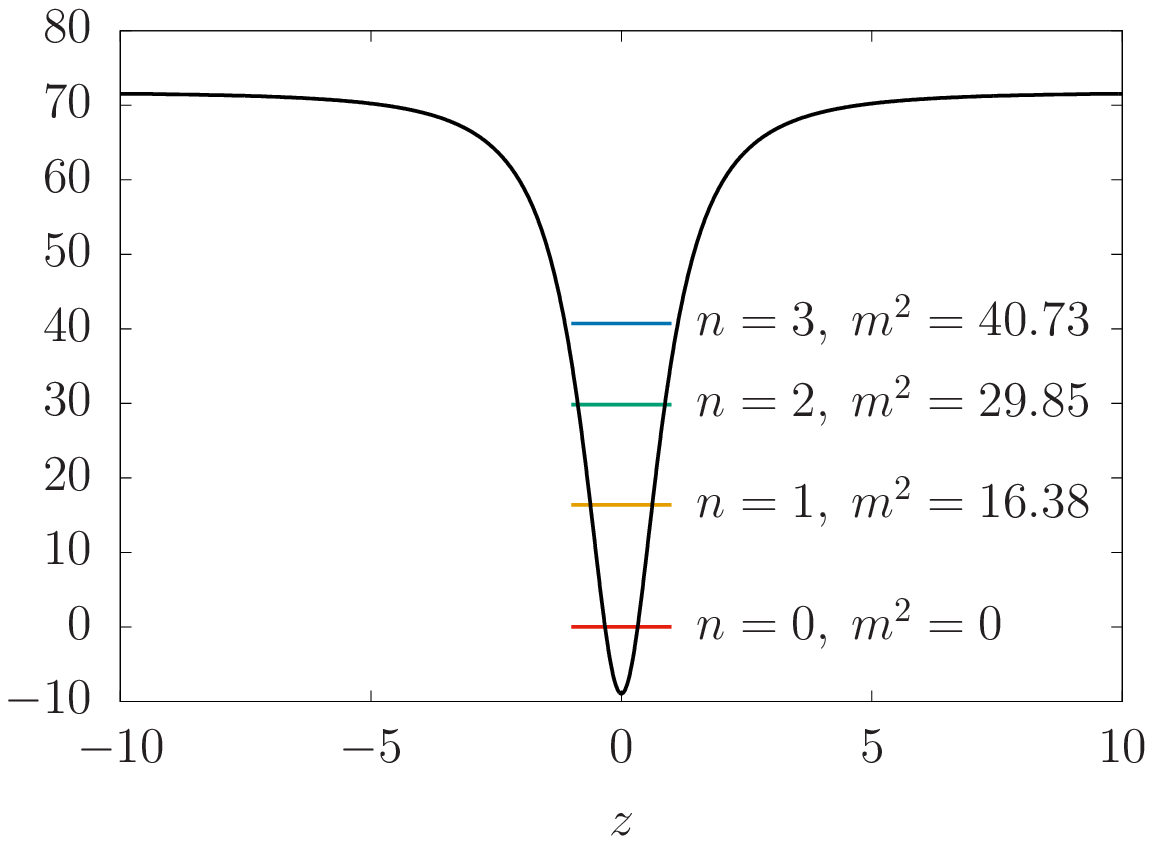}
\includegraphics[width= 0.49\textwidth]{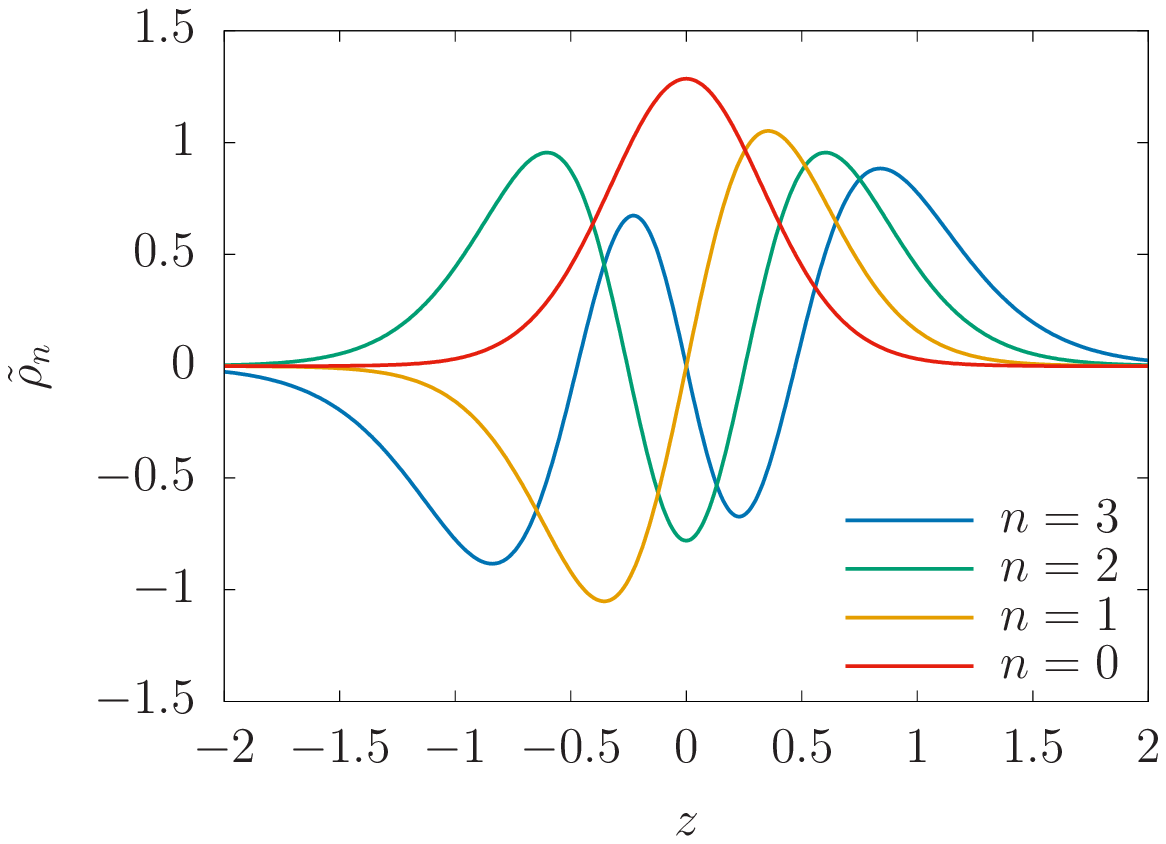}
\caption{In the left figure, {  the} black line represents the potential $V_\mathrm{III}(z)$, and  colored lines represent {   the position of  mass spectra in the potential}. The right figure shows the corresponding solutions of $\tilde{\rho}_{n}$ (nomalized).  Parameters $ {  \mathrm{C}_{3}= 0.5},\; \mathrm{C}_{2}= 20$, and $k=1$. }
\label{figrhozVz1PT}
\end{center}
\end{figure}

\begin{figure}[htbp]
\begin{center}
\includegraphics[width= 0.49\textwidth]{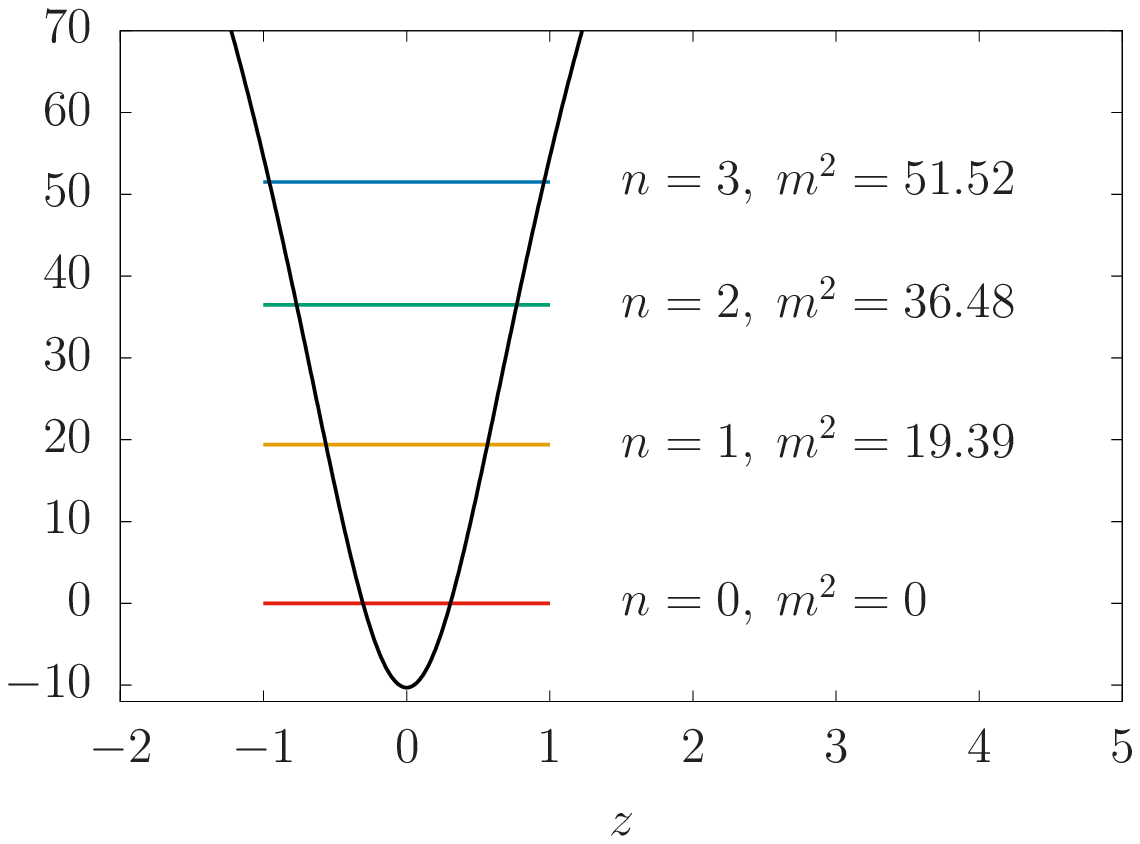}
\includegraphics[width= 0.49\textwidth]{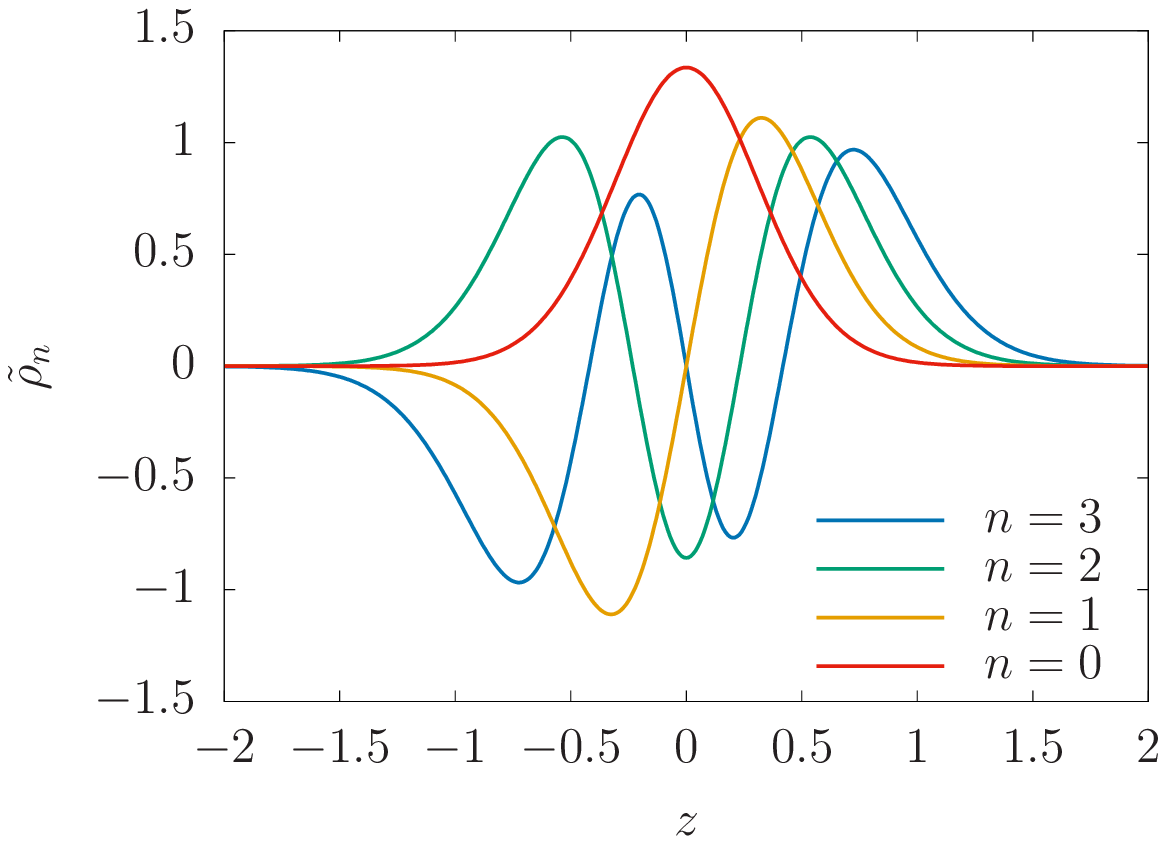}
\caption{In the left figure, {  the} black line represents the potential $V_\mathrm{III}(z)$, and   colored lines represent {  the position of  mass spectra in the potential}. The right figure shows the corresponding solutions of $\tilde{\rho}_{n}$ (nomalized).  Parameters $ {  \mathrm{C}_{3}= 0.6},\; \mathrm{C}_{2}= 20$, and $k=1$. }
\label{figrhozVz12}
\end{center}
\end{figure}

The localized massive mode solutions can be obtained by using of numerical method in the case of $\mathrm{C}_{3}\ge 1/2$. For example we show some massive solutions in Fig. \ref{figrhozVz1PT} with parameters $ {  \mathrm{C}_{3}= 0.5},\; \mathrm{C}_{2}= 20$, and $k=1$.
Another example is shown in Fig. \ref{figrhozVz12} with parameters $\mathrm{C}_{3}= 0.6,\; \mathrm{C}_{2}= 20$, and $k=1$.

\section{Conclusions}\label{Cons}

 In order to localize U(1) gauge vector fields on branes in 5D (asymptotic) AdS$_{5}$ spacetime with infinitely extra-dimension, we  propose a 5D U(1) gauge vector fields action which include  the non-minimal coupling with gravity and satisfy the gauge invariant.  In this work we propose three kinds of coupling, they  all support the  localization of zero mode, but only the  third one \eqref{FR3} support the localization of massive modes. With the third kind of  coupling \eqref{FR3}, there is a parameter $\mathrm{C}_{3}$ which can control the localization of massive modes, namely, when
\begin{itemize}
\item  $\mathrm{C}_{3} = 1/n$, there are finite number of localized massive modes,  
\item  $\mathrm{C}_{3} > 1/n$, there are  infinite number of localized massive modes, 
\item  $ 0 <\mathrm{C}_{3} < {1}/{n},  $ there is not localized massive mode,
\end{itemize}
where  $n$ is a positive number determined by the scalar curvature $R$ \eqref{Rzinf}. Moreover,  our method can exclude the  massive tachyonic modes and can be used   not only in the thin braneword models but also in the thick ones.

\section*{Acknowledgments}

{ 
We thanks Prof. Yu-Xiao Liu for helpful discussion. We also thank the referee for helpful comments and suggestions. This work was supported by the National Natural Science Foundation of China (Grant Nos.  11305095, 11522541, 11375075,  and 11705070), } the Natural Science Foundation of Shandong Province, China (Grant No. ZR2013AQ016), and the Scientific Research Foundation of Shandong University of Science and Technology for Recruited Talents (Grant No. 2013RCJJ026).

%
%
%

\bibliographystyle{JHEP}
\bibliography{C:/yun/library_zhao/articles_all}

\end{document}